# High-pressure isostructural transition in nitrogen

Shan Liu, Meifang Pu, Qiqi Tang, Feng Zhang, Binbin Wu, Li Lei*

*Institute of Atomic and Molecular Physics, Sichuan University, 610065 Chengdu, People's Republic of China*

*Electronic mail: lei@scu.edu.cn*

Understanding high-pressure transitions in prototypical linear diatomic molecules, such as hydrogen[1-3], nitrogen[4-7], and oxygen[8,9], is an important objective in high-pressure physics. Recent ultrahigh-pressure study on hydrogen revealed that there exists a molecular-symmetry-breaking isostructural electronic transition in hydrogen[3]. The pressure-induced symmetry breaking could also lead to a series of solid molecular phases in nitrogen associated with the splitting of vibron[10], walking a path from fluid nitrogen to molecular solids ($\beta$-N$_2$[11,12], $\delta$-N$_2$[5,13,14], $\delta_{loc}$-N$_2$[14,15], $\varepsilon$-N$_2$[5,16], $\zeta$-N$_2$[7,17], $\zeta'$-N$_2$[4], $\kappa$-N$_2$[7] and $\lambda$-N$_2$[18]), and then to polymetric nitrogen phases (cg-N[17,19-21], LP-N[22] and HLP-N[23]). However, isostructural transition in nitrogen has not been reported so far, which may provide crucial information on the crystallographic nature of the pressure-induced dissociation transitions in nitrogen. Here we present a high-pressure Raman scattering and high-pressure angle dispersive X-ray diffraction study of the previously reported $\lambda$-N$_2$, and reveal an isostructural transition from the $\lambda$-N$_2$ to a new derivative phase, $\lambda'$-N$_2$. Under compression,

**$\lambda$-N$_2$ remain in the monoclinic crystal lattice, accompanied by a monotonic increase in anisotropy. The pressure-dependent decrease of the unit cell parameters exhibits a discontinuity at the pressure of 54 GPa, accompany with sudden broadening and turnover of vibron.**

The $\lambda$-N$_2$ has been reported as a low-temperature, high-pressure (LTHP) molecular solid phase, which was first synthesized by compression of high purity liquid nitrogen at 77 K from ambient pressure to above 32 GPa[18]. The $\lambda$-N$_2$ adopts a monoclinic structure (space group $P2_1/c$) with unit cell parameter $a$= 2.922 Å, $b$= 2.8o91 Å, $c$= 5.588 Å, $\beta$= 132.54 at 40 GPa[24]. The $P2_1/c$ structure is based on an fcc symmetry, but with molecular displacements and a significant cell distortion[24]. Like other solid nitrogen phases, $\lambda$-N$_2$ possesses combined characteristics of crystalline solid and diatomic molecules, exhibiting distinct three Raman vibrons in low-frequency region and four Raman phonons in high-frequency region[18]. The $\lambda$-N$_2$ exhibits an exceptionally wide range of pressure-temperature ($P$-$T$) stability, covering $P$-$T$ space of nine other ordinary phases, such as $\varepsilon$, $\zeta$, $\zeta'$ $\kappa$, $\iota$, and $\theta$[18]. Previous Raman scattering and powder x-ray diffraction data do not show any phase boundary over the wide region of $P$-$T$ space, in which at least six solid phases of molecular nitrogen are all observed. So, it is necessary to evaluate the formation conditions, phase relations and $P$-$V$ curve of the previously reported $\lambda$-N$_2$.

High pressure was generated using a diamond-anvil cell (DAC) with 100~300 μm culets. Rhenium was used as the gasket material precompressed to 20-30 μm thickness with a sample chamber drilled using laser cutting to produce a 30-100 μm diameter holes. In the past four years, $\lambda$-N$_2$ and its isostructural derivative $\lambda'$-N$_2$ have been successfully produced in seven separate experiments by compression of high purity liquid nitrogen at low temperature from ambient pressure. Five typical *P-T* paths via compression at 77 K up to at least 30, 37, 79, 98, 107 GPa are shown in the Fig. 1a. The obtained $\lambda$-N$_2$ ($\lambda'$-N$_2$) nitrogen sample was continually compressed up to 150 GPa at room temperature. High-pressure Raman scattering experiments were carried out on a custom-built confocal Raman spectrometry system in the back-scattering geometry based on Andor Shamrock triple grating monochromator with an attached ANDOR Newton EMCCD, excitation by a solid-state laser at 532 nm, and collection by a 20× Mitutoyo objective. In high-pressure Raman experiments, the pressure was jointly monitored through the high-frequency edge of the diamond phonon[25,26] and the ruby fluorescence[27]. The ruby pressure scale is typically 1-3 GPa (<70 GPa) lower than the diamond Raman scale in our experiments. Considering ruby fluorescence method for the high pressure above 100 GPa cannot be read accurately, we only use the diamond phonon as pressure calibration in the high-pressure Raman experiments in this work.

The high-pressure angle dispersive X-ray diffraction (ADXRD) data were collected at pressures from 35 GPa to 61 GPa at the 4W2 beam line of the Beijing Synchrotron Radiation Facility (BSRF, China) with pressure measured by the ruby fluorescence method. A Si (111) monochromator was used to tune the synchrotron source with a wavelength of 0.6199 Å. The incident X-ray beam was focused to approximately 26×8 μm$^2$ full width at half maximum (FWHM) spot by a pair of Kirkpatrick-Baez mirrors. The two-dimensional diffraction patterns were recorded by a Mar345 image plate detector and analyzed with the program Fit2D[28].

$\lambda$-N$_2$ exhibits an exceptionally wide range of *P-T* stability[18]. In the present work, five different *P-T* paths, which would produce $\lambda$-N$_2$, are used to probe the formation condition and the phase relations (Fig. 1a). Fig. 1b shows the room-temperature Raman spectra of the as-obtained phases, which forms on compression from ambient pressure at low temperatures via the five paths. The Raman spectroscopy of $\lambda$-N$_2$ was first characterized by Frost, *et al.*[18], and it exhibits different behaviors in comparison with the ordinary phases obtained via regular compression path (see supplemental material). It is not difficult to distinguish the $\lambda$-N$_2$ phase from those ordinary high-pressure phases obtained via regular compression path due to the distinct Raman bands of the $\lambda$-N$_2$. The room-temperature transition pressure from $\lambda$-N$_2$ to $\varepsilon$-N$_2$ is about 30 GPa (see path I in Fig. 1b), which is slightly lower than the previously reported pressure (32 GPa) in Ref. 18. If we use ruby scale gives lower transition pressure (~ 28 GPa). We find that the $\lambda$-N$_2$ undergoes an isostructural phase transition to a new $\lambda'$-N$_2$ at about 54 GPa under room-temperature compression, and we will discuss in details in the following section. The former *P-T* space of $\lambda$-N$_2$ in Ref. 18 should be divided into two parts, a lower-pressure region for the $\lambda$-N$_2$ and a higher-pressure region for the $\lambda'$-N$_2$ (Fig. 1a). A coexistence phase of $\lambda'$-N$_2$ and $\zeta$-N$_2$ is observed at 107 GPa (see path V in Fig. 1b), therefore the low-temperature boundary between $\lambda'$- and $\zeta$-N$_2$ should be at 107 GPa, which is much lower than the previously reported 140 GPa in the Ref. 18. We also notice that there was no studied path over the range of pressure above 100 GPa at 77 K in the previous study[18]. In the present work, the obtained $\lambda$-N$_2$ ($\lambda$'-N$_2$) samples were continually compressed up to 150 GPa at room temperature (Fig. 1c). The samples were transparent at lower pressure (< 120 GPa), becomes dull red at about 130 GPa, and almost opaque at about 145 GPa (Fig. 1a). At 150 GPa, the Raman bands become broadened and weak, but they still can be clearly observed (Fig. 1c). It is possible for the $\lambda$'-N$_2$ entering the Raman-inactive amorphous $\eta$-N$_2$ above 150 GPa, revealing that the room-temperature transition pressure from $\lambda$'-N$_2$ to $\eta$-N$_2$ should be extended to at least 150 GPa instead of the previously reported 140 GPa[18], as depicted in the Fig. 1a.

The $\lambda$-N$_2$ undergoes a high-pressure isostructural phase transition to a new $\lambda$'-N$_2$, the transition is strongly supported by both high-pressure Raman scattering and high-pressure X-ray diffraction experiments. As shown in the Fig. 1c, seven Raman modes can be clearly observed for the $\lambda$-N$_2$, four of which are low-frequency lattice phonons, and three of which are high-frequency vibrons. According to the reported space group $P2_1/c$ for the $\lambda$-N$_2$[24], six Raman active optical modes come from Brillouin zone center, $\Gamma = 3A_g + 3B_g$. These six Raman modes can be assigned as $A_g^{(1)}$, $A_g^{(2)}$, $A_g^{(3)}$, $B_g^{(1)}$, $B_g^{(2)}$ and $B_g^{(3)}$. Thereinto, $A_g^{(1)}$, $A_g^{(2)}$, $B_g^{(1)}$ and $B_g^{(2)}$ are indexed as low-frequency lattice phonon modes, $A_g^{(3)}$ and $B_g^{(3)}$ are indexed as high-frequency vibron modes. Vibration characteristic analysis shows that the vibron modes of solid molecular nitrogen are associated with linear symmetrical stretching vibrations in the nitrogen diatomic molecules[10]. In the case of the $\lambda$-N$_2$, each unit cell contains two types of symmetrical polarization nitrogen molecular pairs (see Fig. 2a), one lies in the center of the unit cell (blue molecules), and the other (red molecules) locates in the edges of the unit cell. The $A_g^{(3)}$ and $B_g^{(3)}$ vibron modes, therefore, are associated with the linear symmetrical stretching vibrations of the red and blue diatomic molecular pairs, respectively (Fig. 1d).

A weak high-frequency mode, $A_g'$, can be also observed under room-temperature compression, which is also reported by Frost *et al.*[18] $A_g'$ mode gradually approaches to the $A_g^{(3)}$ mode with increasing pressure, and eventually merges with the $A_g^{(3)}$ mode above 100 GPa (Fig. 1c). We note that the $A_g'$ mode cannot be observed under low-temperature (77 K) compression[18], and it is also very weak at relatively lower pressures (< 80 GPa), but becomes relatively stronger at higher pressures (> 90 GPa, < 130 GPa), as shown in the Fig. 1c. Although group theory analysis indicates that there are only two high-frequency vibron modes, $A_g^{(3)}$ and $B_g^{(3)}$, for the $\lambda$-$N_2$ with $P2_1/c$, the pressure-induced local symmetry breaking may lead to the emergence of $A_g'$ mode. For a perfect unit cell of $\lambda$-$N_2$ molecular solid depicted in the Fig. 1d, two red diatomic molecular pairs in the edges of unit cell are parallel to each other, but the orientation stability of the polarized molecular pairs are easy to be damaged at high pressure (> 30 GPa) and at higher temperature (300 K) due to the long *c*-axis in the unit cell (*c*/*a*~1.87 at 35 GPa) and the low bulk modulus (*B*~68 GPa at 35 GPa) for the $\lambda$-$N_2$. The local structural symmetry is broken by high stress, and the 'perfect' structure are likely distorted under extreme stain environment, which is characterized by the distances between the red molecule pairs and the blue molecule pairs are no longer a fixed value, and the polarization directions of these red molecular pairs are no longer accordant. The observed $A_g'$, therefore, could be indexed as a local symmetry breaking mode (LSBM), which is attributed to the structural instability and local molecular polarization property of $\lambda$-$N_2$ induced by high pressure. The intensity of the LSBM, of cause, increases with the degree of disorder. As shown in Fig. 1c, the $A_g'$ mode becomes stronger at higher pressure, and also decays with the phase transformation to the $\eta$-$N_2$.

The pressure dependence of Raman modes of $\lambda$-$N_2$ ($\lambda'$-$N_2$) is shown in Fig. 1e, the previous $\lambda$-$N_2$ data from Ref [18] and the regular compression route from Ref. 10 are also shown for comparison. The extensive data in this work cover the pressure range from 30 to 150 GPa, which fills the evolutionary gap of the low-frequency lattice modes between 70 to 110 GPa in Ref. 18. Our data are in good agreement with the previous data[18] at lower pressure (<80 GPa); while there is an increasing discrepancy for the pressures higher than 80 GPa. The discrepancy cannot be attributed to the different pressure scales employed in the high-pressure experiments. In the present study, the pressure was measured by both the diamond Raman method[25, 26] and the ruby fluorescence method[27]. The discrepancy between these two methods is smaller than 3 GPa below 100 GPa.

The low-frequency Raman lattice modes, which are associated with intermolecular interactions, increase with increasing pressure; while the high-frequency Raman vibron modes, which are associated with intramolecular interactions, initially increase with pressure, but this trend stop at a certain pressure, above which the vibron frequency decreases with increasing pressure. Fig. 1e shows that the vibron frequency of $A_g^{(3)}$ mode reaches its maximum (ca. 2393 cm$^{-1}$) at about 50 GPa, and then turns over and gradually decreases with further increasing pressure. At 150 GPa, the vibron frequency of $A_g^{(3)}$ decreases to ca. 2333 cm$^{-1}$, which is very close to the original frequency (ca. 2332 cm$^{-1}$) at ambient pressure.

As shown in Fig. 1f, most of the vibrons experience the frequency turnover, and the pressure derivation of the $A_g^{(3)}$ mode gets the zero at 54 GPa, and the width of $A_g^{(3)}$ mode broadens considerably with pressure and exhibits discontinuity at 55.8 GPa. The sudden broadening and turnover of vibron imply the weakening and reforming of the intramolecular interactions and likely contribute to a first-order transition in molecular solid under high pressure. In the case of ordinary phases via regular room-temperature compression path, the dominated vibron, $v_{2c}$, turns over at about 84 GPa (Fig. 1f), at which a phase transition from $\zeta$- to $\zeta'$-N$_2$ occurs[10]. $\zeta'$-N$_2$ is a derivative phase of $\zeta$-N$_2$. Although the exact structure of $\zeta'$-N$_2$ need to be further determined, it could be ascribed to the slight rotation or distortion of molecular clusters in the $\zeta$-N$_2$ structure[10]. In the case of phase transition in deuterium, the Raman vibron frequency first increases with pressure, and then turns over at 62 GPa and 77 K, at which the transition to the breading symmetry phase (BSP)[1,2]. Therefore, the turnover of the Raman main peak $A_g^{(3)}$ mode implies a phase transition from $\lambda$-N$_2$ to a new nitrogen phase $\lambda'$-N$_2$.

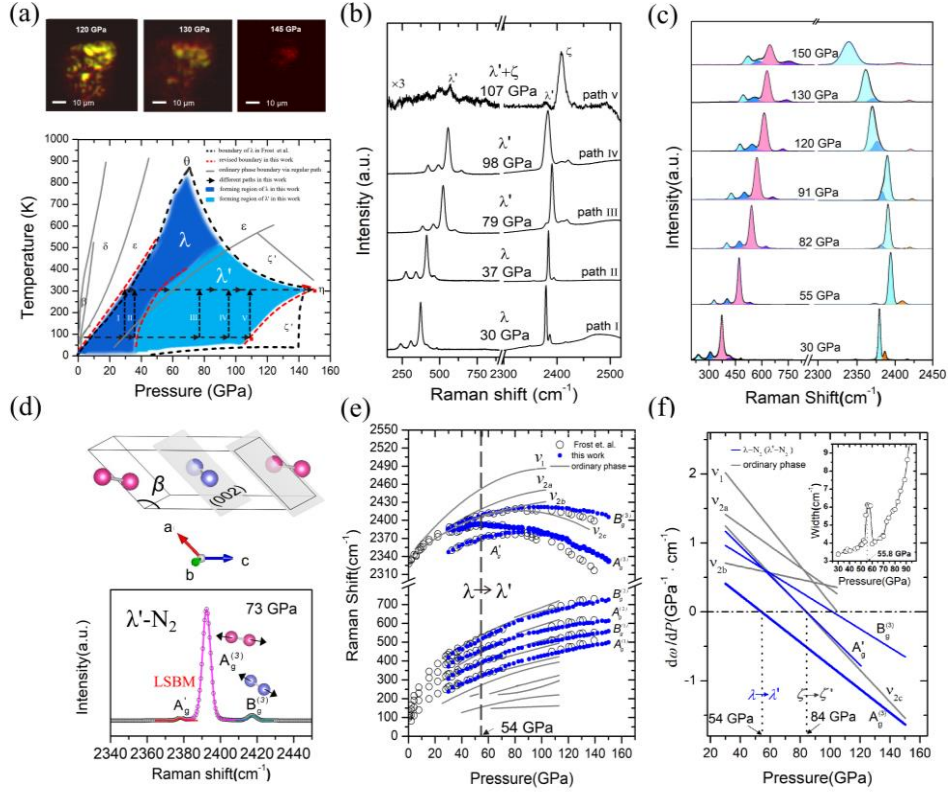

Fig. 1 (a) Top: Microscope image of the nitrogen samples at 120, 130 and 145 GPa and room temperature. These photos were made with the same transmitting light. Bottom: Transformation and phase diagram of nitrogen. The deep blue section and light blue section are the forming region of $\lambda$-$N_2$ and $\lambda$'-$N_2$, respectively. The black dashed lines show the boundaries of $\lambda$-$N_2$ in Ref. 18, the red dashed lines represent the revised transition boundaries of $\lambda$-$N_2$ and $\lambda$'-$N_2$ in this work, the black dashed arrows indicate five *P-T* paths which produce $\lambda$-$N_2$ and/or $\lambda$'-$N_2$. The gray lines show the phase boundaries of nitrogen previously observed under compression at room temperature. (b) The room-temperature Raman spectra of the as-obtained samples produced by the five key paths. (c) Representative high-pressure Raman spectra of $\lambda$-$N_2$ and $\lambda$'-$N_2$ on room-temperature compression up to 150 GPa. Black arrow indicates the local symmetry breaking mode $A_g$'. (d)Top: crystal structures for $\lambda$-$N_2$ with space group *P*2$_1$/*c*; Bottom: the vibrational assignment of the high-frequency Raman vibrons. € Pressure dependence of Raman shift of the $N_2$ vibron and phonon. The present data (blue solid circles), the previously reported $\lambda$-$N_2$ data from Ref 18 (gray open circles), and the ordinary nitrogen data obtained via the regular compression path (gray lines)[10]. The larger blue solid circles represent the main Raman peak ($A_g^{(3)}$) of $\lambda$-$N_2$ and $\lambda$'-$N_2$. (f) The pressure derivation of Raman vibron frequency as a function of pressure, the blue lines for the $\lambda$-$N_2$ ($\lambda$'-$N_2$), and gray lines for the ordinary phases via regular path. Dashed lines indicate the turnover of vibron at the phase transition pressure. Inset is the pressure dependence of the widths (FWHM) of $A_g^{(3)}$ mode at room temperature.

The isostructural phase transition from $\lambda$-$N_2$ to $\lambda'$-$N_2$ is also supported by our high-pressure X-ray diffraction experiments. ADXRD data on a $\lambda$-$N_2$ nitrogen sample at pressures from 35 GPa to 61 GPa provides another evidence for the proposed isostructural phase transition between $\lambda$-$N_2$ and $\lambda'$-$N_2$ (Fig. 2a). In the XRD patterns, there is an overlapping of peaks between the nitrogen and the Re gasket. Because of the preferred orientation of the solid molecular phase under pressure, peak along the ($h$00) direction could be discerned. A total of five diffraction peaks can be followed from 35 GPa up to 61 GPa. Le Bail fits of the present data give a good fit to the reported monoclinic structure with $P2_1/c$ space group, and remain isostructural except for a discontinuity in unit cell parameters. The following unit cell parameters could be obtained: $a$=2.9605(5) Å, $b$=2.9760(8) Å, $c$=5.5376(9) Å ($V$=35.45084 Å$^3$) with $\beta$=133.4(5)° at 35.7 GPa, which is in a little bit difference in Ref 18, the unit cell parameters $a$ = 3.051(7) Å, $b$ = 3.066(5) Å, $c$ = 5.705(13) Å ($V$=35.49599 Å$^3$) with $\beta$=131.65(5)° at 34 GPa.

There is an obvious discontinuity at about 51 GPa in the pressure dependence of the unit cell parameters in terms of diffraction planes (Fig. 2b), the unit cell axial ratios (Fig. 2c), and the $\beta$ angle (Fig. 2d), suggesting the occurrence of an isostructural phase transition at this pressure point. It is worth noting that the (002) plane, which associates with the intermolecular distance between blue molecular pairs and red molecular pairs (see Fig. 1d), exhibits significantly abnormal compressive behaviors about 51 GPa (Fig. 2b). The unit cell axial ratios keep decreasing and the anisotropy continue to grow under compression below 51 GPa, however, the obvious collapses of axial ratios are observed at about 51 GPa (Fig. 2c). The transition is not caused by major crystallographic changes of the monoclinic structure. The refined unit cell parameters for $\lambda'$-N$_2$ are $a$=2.8505(9) Å, $b$=2.8283(2) Å, $c$=5.4071(9) Å ($V$=31.98705 Å$^3$) with $\beta$=132.8(5)° at 51.1 GPa. The $\beta$ angle first decreases with increasing pressure and then sudden increase at about 51 GPa (Fig. 2d), which coincides with the abrupt changes in axis ratios. The $\beta$ angle is also a key parameter for the isostructural transition, associated with the structural instability of molecular solid and orientation of the local molecular pairs. In the present high-pressure XRD experiments, the pressure was only measured by the ruby fluorescence method. We find that the pressure scaled by the ruby fluorescence is 1-3 GPa lower than the pressure determined by the diamond Raman method. So, it is not difficult to understand the transition pressure (51 GPa) determined by the high-pressure XRD experiments is slightly lower than the pressure (54 GPa) determined by the high-pressure Raman experiments.

The $P$-$V$ compression curves for $\lambda$-N$_2$[18,24,29], $\lambda'$-N$_2$, $\delta$-N$_2$[5], $\delta_{loc}$-N$_2$[14,15], $\varepsilon$-N$_2$[5,6], $\zeta$-N$_2$[6], $\kappa$-N$_2$[7], $\eta$-N$_2$[7], and cg-N[6,7] are plotted in Fig. 2e. In this work, high-pressure ADXRD experiments on $\delta$-N$_2$, $\delta_{loc}$-N$_2$, $\varepsilon$-N$_2$ have also been performed, and the obtained the pressure-volume relation of the three phases (see supplemental material) are also shown in the Fig. 2e. The data are in in good agreement with the previous work[5,6]. The volume collapse at 60 GPa indicates the transition from $\varepsilon$-N$_2$ to $\zeta$-N$_2$[6], The compression curves of $\lambda$-N$_2$ and $\lambda'$-N$_2$ are very near, but discontinuous, underlining the isomorphic structure between these two phases. Actually, a discontinuity at about 60 GPa in the *P-V* curve was also found in the Ref. 18. The present experimental data for the λ-N$_2$ are in good agreement with the calculation result in Ref. 24, 29. However, the experimental volumes measured for the λ-N$_2$, reported in Ref. 18, are much higher than the calculations and our data. Our *in-situ* XRD measurements confirm the previous observation of sudden broadening and turnover of vibron during the isostructural transition.

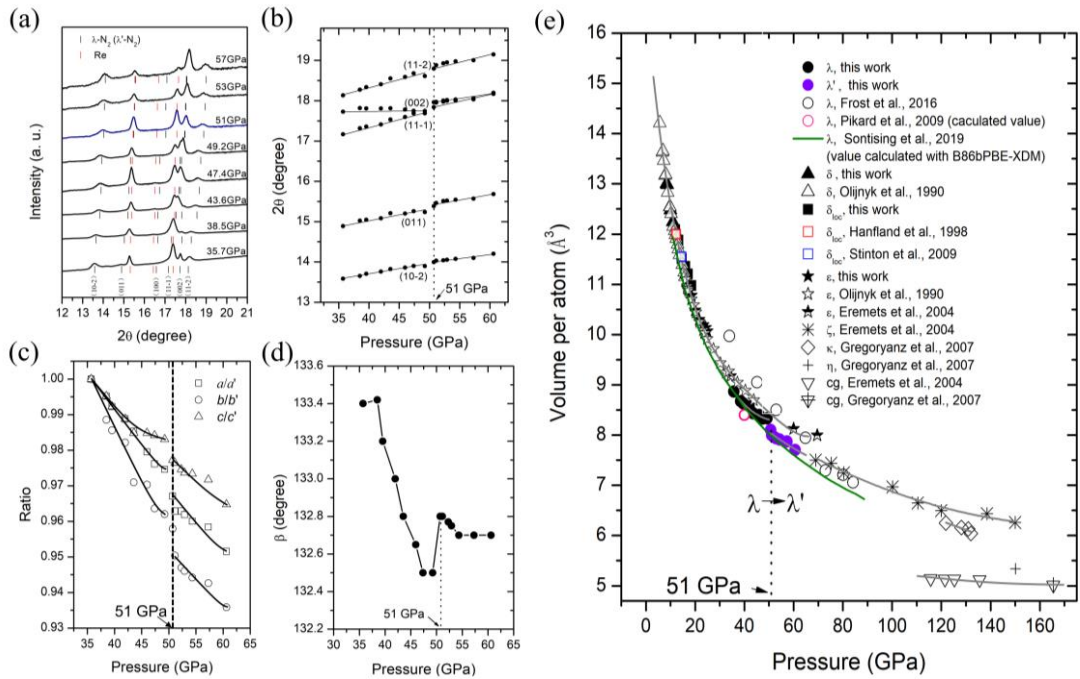

Fig. 2 (a) Representative high-pressure XRD patterns of λ-N$_2$ and λ'-N$_2$. Black tick lines show positions of the λ-N$_2$ and λ'-N$_2$ phase and red tick lines indicate the Re gasket. Diffraction planes of λ-N$_2$ (λ'-N$_2$) are indexed (b) Bragg angles of the diffraction planes as a function of pressure. (c) Evolution of the unit cell axial ratios (*a/a'*, *b/b'*, *c/c'*) with pressure, therein *a'*, *b'* and *c'* are the unit cell parameters under 35.7 GPa at room temperature. (d) Evolution of the *β* angle of the monoclinic cell with pressure. € Volume per nitrogen atom vs pressure. Solid shapes represent the present experimental data, open red circle is calculation data for λ-N$_2$ from Ref. 24, open black circle is the experimental data for λ-N$_2$ from Ref. 18. Experimental data for ordinary phases via regular path are from Ref. 5-7, 14, 15. Gray lines are fits of experimental data, and green line is calculation curve from Ref. 29. Dashed line indicates the discontinuity at about 51 GPa.

In summary, the formation conditions, phase relations and *P-V* curve of the previously reported $\lambda$-$N_2$ have been carefully evaluated, and a high-pressure isostructural phase transition with changes of vibrational properties and lattice parameters from $\lambda$-$N_2$ to a new $\lambda'$-$N_2$ was observed by both Raman scattering and X-ray diffraction experiments. Since the discovery of isostructural phase transition in hydrogen, such transition was also observed in nitrogen, which sheds light on the properties of molecular solid under high pressure and the pressure-induced dissociative transition in linear diatomic molecular.

## ACKNOWLEDGMENTS

We acknowledge support by the National Natural Science Foundation of China (Grant No. 11774247), Chinese Academy of Sciences (Grant No. 2017-BEPC-PT-000568, 2019-BEPC-PT-003237), and the Science Foundation for Excellent Youth Scholars of Sichuan University (Grant No. 2015SCU04A04).

# References

1. Silvera, I. F., & Wijngaarden, R. J. New low-temperature phase of molecular deuterium at ultrahigh pressure. *Phys. Rev. Lett.*, **47**, 39 (1981).

2. Baer, B. J., Evans, W. J., & Yoo, C. S. Coherent anti-stokes Raman spectroscopy of highly compressed solid deuterium at 300 K: Evidence for a new phase and implications for the band gap. *Phys. Rev. Lett.,* **98**, 235503 (2007).

3. Ji, C., et al. Ultrahigh-pressure isostructural electronic transitions in hydrogen. *Nature*, **573**, 558-562 (2019).

Gregoryanz, E., Goncharov, A. F., Hemley, R. J., & Mao, H. K. High-pressure amorphous nitrogen. *Phys. Rev. B,* **64**, 052103 (2001)

5. Olijnyk, H. High pressure x-ray diffraction studies on solid N2 up to 43.9 GPa. *J. Chem. Phys.,* **93**, 8968-8972(1990).

6. Eremets, M. I., et al. Structural transformation of molecular nitrogen to a single-bonded atomic state at high pressures. *J. Chem. Phys.,* **121**, 11296-11300 (2004).

7. Gregoryanz, E., et al. High P-T transformations of nitrogen to 170 GPa. *J. Chem. Phys.,* **126**, 184505 (2007).

8. Akahama, Y., Kawamura, H., Häusermann, D., Hanfland, M., & Shimomura, O. New high-pressure structural transition of oxygen at 96 GPa associated with metallization in a molecular solid. *Phys. Rev. Lett.*, **74**, 4690 (1995).

9. Desgreniers, S., Vohra, Y. K., & Ruoff, A. L. Optical response of very high density solid oxygen to 132 GPa. *J. Phys. Chem.,* **94**, 1117-1122 (1990).

10. Pu, M., et al. Raman study of pressure-induced dissociative transitions in nitrogen. *Solid State Commun.,* 113645 (2019).

11. Streib, W. E., Jordan, T. H., & Lipscomb, W. N. Single-Crystal X-Ray Diffraction Study of β Nitrogen. *J. Chem. Phys.*, **37**, 2962-2965 (1962).

12. Schiferl, D., Cromer, D. T., Ryan, R. R., Larson, A. C., LeSar, R., & Mills, R. L. Structure of $N_2$ at 2.94 GPa and 300 K. *Acta Crystallogr. C,* **39,** 1151-1153(1983).

13. Cromer, D. T., Mills, R. L., Schiferi, D., & Schwalbe, L. A. The structure of $N_2$ at

49 kbar and 299 K. *Acta Crystallogr. B*, **37,** 8-11 (1981).

14. Stinton, G. W., Loa, I., Lundegaard, L. F., & McMahon, M. I. The crystal structures of δ and δ* nitrogen. *J. Chem. Phys.,* **131,** 104511 (2009).

15. Hanfland, M., Lorenzen, M., Wassilew-Reul, C., & Zontone, F. Structures of molecular nitrogen at high pressures. *Rev. High Pressure Sci. Technol.,* **7,** 787-789 (1998).

16. Mills, R. L., Olinger, B., & Cromer, D. T. Structures and phase diagrams of $N_2$ and CO to 13 GPa by x-ray diffraction. *J. Chem. Phys.,* **84,** 2837-2845 (1986).

17. Bini, R., Ulivi, L., Kreutz, J., & Jodl, H. J. High-pressure phases of solid nitrogen by Raman and infrared spectroscopy. *J. Chem. Phys.,* **112,** 8522-8529 (2000).

18. Frost, M., Howie, R. T., Dalladay-Simpson, P., Goncharov, A. F., & Gregoryanz, E. Novel high-pressure nitrogen phase formed by compression at low temperature. *Phys. Rev. B,* **93**, 024113 (2016).

19. Eremets, M. I., Gavriliuk, A. G., Trojan, I. A., Dzivenko, D. A., & Boehler, R. Single-bonded cubic form of nitrogen. *Nat. Mater.*, **3**, 558 (2004).

20. Lipp, M. J., et al. Transformation of molecular nitrogen to nonmolecular phases at megabar pressures by direct laser heating. *Phys. Rev. B*, **76,** 014113 (2007).

21. Eremets, M. I., Gavriliuk, A. G., & Trojan, I. A. Single-crystalline polymeric nitrogen. *Appl. Phys. Lett.,* **90,** 171904 (2007).

22. Tomasino, D., Kim, M., Smith, J., & Yoo, C. S. Pressure-induced symmetry-lowering transition in dense nitrogen to layered polymeric nitrogen (LP-N) with colossal Raman intensity. *Phys. Rev. Lett.,* **113,** 205502 (2014).

23. Laniel, D., Geneste, G., Weck, G., Mezouar, M., & Loubeyre, P. Hexagonal Layered Polymeric Nitrogen Phase Synthesized near 250 GPa. *Phys. Rev. Lett.*, **122,** 066001 (2019).

24. Pickard, C. J., & Needs, R. J. High-pressure phases of nitrogen. *Phys. Rev. Lett.,* **102**, 125702 (2009).

25. Akahama, Y., & Kawamura, H. Pressure calibration of diamond anvil Raman gauge to 310 GPa. *J. Appl. Phys.,* **100**, 043516 (2006).

26. Dalladay-Simpson, P., Howie, R. T., & Gregoryanz, E. Evidence for a new phase

of dense hydrogen above 325 gigapascals. *Nature.* **63**, 529 (2016).

27. Mao, H. K., Xu, J. A., & Bell, P. M. Calibration of the ruby pressure gauge to 800 kbar under quasi-hydrostatic conditions. *J. Geophys. Res.,* **91**, 4673-4676 (1986).

28. Hammersley, A. P., Svensson, S. O., Hanfland, M., Fitch, A. N., & Hausermann, D. (1996). Two-dimensional detector software: from real detector to idealised image or two-theta scan. *High Press. Res.,* **14**, 235 (1996).

29. Sontising, W., & Beran, G. J. Theoretical assessment of the structure and stability of the λ phase of nitrogen. *Phys. Rev. Materials*, **3**, 095002 (2019).

Supplementary Materials

# High-pressure isostructural transition in nitrogen

Shan Liu, Meifang Pu, Qiqi Tang, Feng Zhang, Binbin Wu, Li Lei[*]

*Institute of Atomic and Molecular Physics, Sichuan University, 610065 Chengdu, People's Republic of China*

*Electronic mail: lei@scu.edu.cn

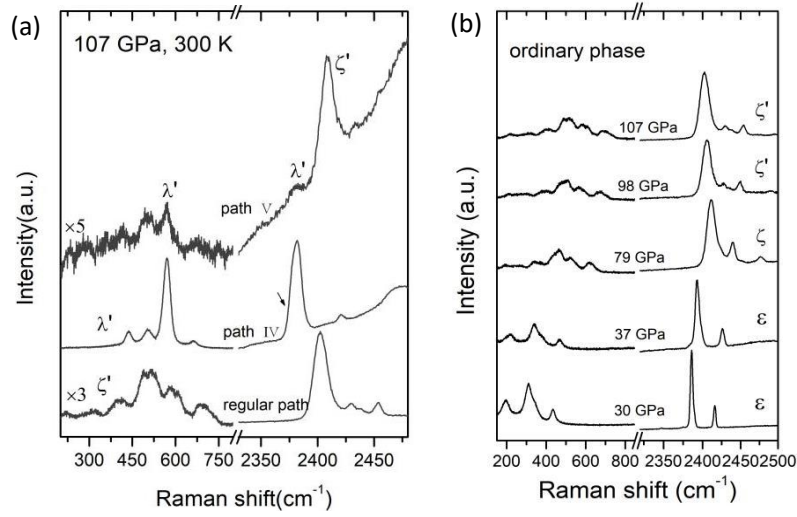

Fig. S1 (a) Raman spectra of path V at 107 GPa and 300 K in compression with pure $\lambda$' and pure $\zeta$ phases, a coexistence phase of $\lambda$'-$N_2$ and $\zeta$-$N_2$ is observed at 107 GPa and 300 K. (b) Raman spectra of ordinary nitrogen phases via regular path at some typical pressures.

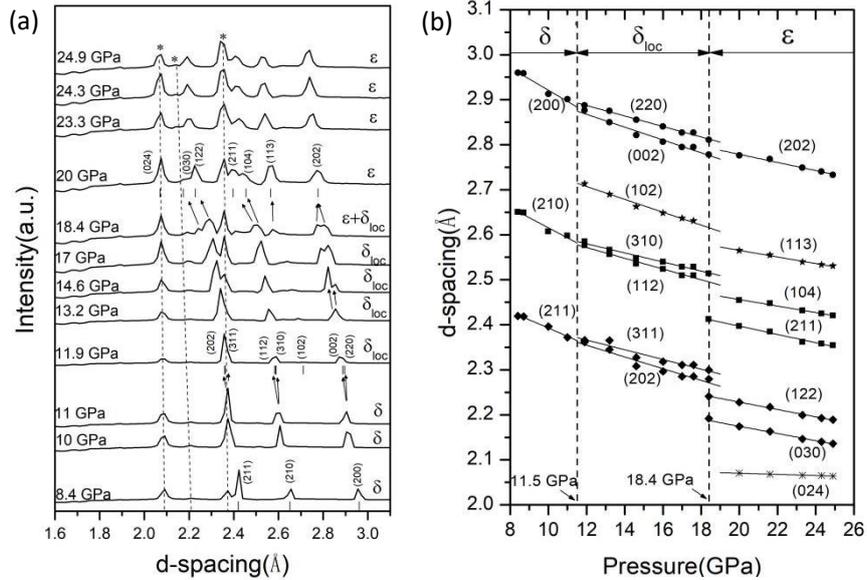

Fig. S2 (a) Integrated x-ray diffraction patterns of ordinary nitrogen phases collected at room temperature from 8.4 to 24.9 GPa at Beijing Synchrotron Radiation Facility (BSRF, China) with an x-ray wavelength of $\lambda$ = 0.6199 Å. The black dashed lines indicate the three diffraction lines of Re gasket, also identified at 24.9 GPa by black asterisks, which could be followed with pressure. Vertical bars indicated the $d$ spacing positions. Nitrogen from 8.4 GPa to 11 GPa was identified as the $\delta$-$N_2$ phase. The onset of $\delta_{loc}$-$N_2$ phase occurs at 11.5 GPa. The phase transition from $\delta_{loc}$ to $\varepsilon$ occurs at 18.4 GPa. (b) The $d$ spacings of the $\delta$, $\delta_{loc}$, and $\varepsilon$ phases as a function of pressure. The black dashed lines indicate the pressure boundary between these phases.

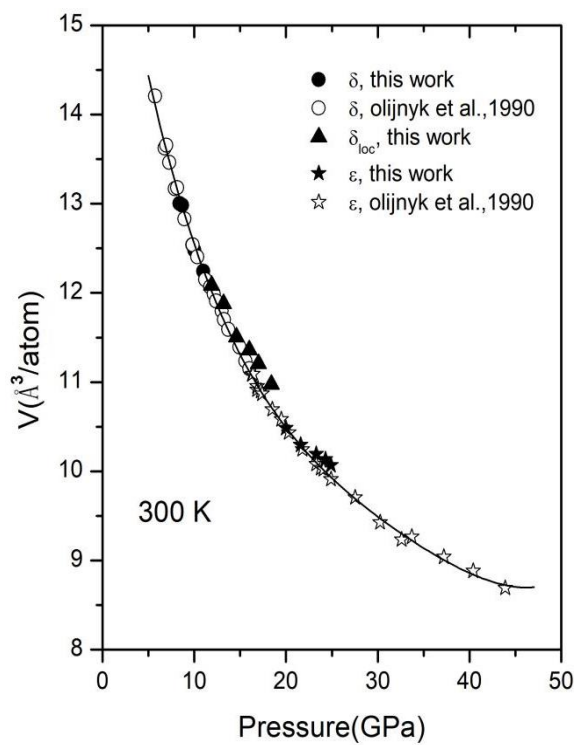

Fig. S3 Pressure-volume relation of $\delta$-$N_2$, $\delta_{loc}$-$N_2$, and $\varepsilon$-$N_2$. Solid shapes are the present experimental data. Open circle and open star are the experimental data from Ref. [5] for $\delta$-$N_2$ and $\varepsilon$-$N_2$, respectively.